\pdfoutput=1 

\documentclass[11pt]{article}
\usepackage[margin=3cm]{geometry}
\usepackage{amsfonts,amsmath,amssymb,amsthm} 
\usepackage{times}
\usepackage{enumerate}
\usepackage{framed}   
\usepackage{graphicx}
\usepackage{subcaption}
\usepackage{verbatim}
\usepackage[bookmarks=false,colorlinks=true,urlcolor=blue]{hyperref}
\usepackage[affil-sl]{authblk}  
\usepackage{url}
\usepackage[normalem]{ulem} 
\usepackage{listings}
\usepackage[inline, shortlabels]{enumitem}   
\usepackage{cite}  

\newtheorem{thm}{Theorem}

\newtheorem{cor}{Corollary}

\newtheorem{prf}{Proof}

\newtheorem{rem}{Remark}
\newtheorem{exam}{Example}

\title{P = FS: Parallel is Just Fast Serial}
\author{Neil J. Gunther}
\affil{\small Performance Dynamics, Castro Valley, California, USA }
\date{}

\begin{document}
\maketitle
\thispagestyle{empty}

\begin{abstract}
We prove that parallel processing with homogeneous processors is logically equivalent to fast serial processing.    
The reverse proposition can also be used to identify obscure opportunities for applying parallelism. 
To our knowledge, this theorem has not been previously reported in the queueing theory literature. 
A plausible explanation is offered for why this might be.  
The basic homogeneous theorem is also extended to optimizing the latency of heterogenous parallel arrays. 
\end{abstract}

\section{Introduction}    \label{sec:intro}
Conventional wisdom holds that parallel architectures have superior performance compared to serial systems. 
Indeed, parallel execution times are generally shorter than serial execution times provided the workload lends itself to the 
necessary partitioning, e.g., threading~\cite{pdcs,mit,speedup} 
What is unrecognized and surprising is that there is a certain correspondence between parallel and serial performance. 
In this note, we show that a {\em parallel} array (P) of $m$ queues  
has the same mean residence time as 
a tandem arrangement of $m$ queues with faster servers, i.e., a {\em fast serial} configuration (FS).   
As far as we are aware, this P = FS observation in Theorem \eqref{thm:homoq} has not been discussed 
in either the queueing theory literature or textbooks (see e.g.,~\cite{qsp,kleinrock,kleinrock2,harrison}), and was first identified and applied by the author~\cite{ppa}.

When it comes to practical application, e.g., cloud computing performances models, 
it can be more useful to employ P = FS in reverse:  
any sequential flow of requests through a tandem queue configuration 
can be replaced by the corresponding array of parallel queues while maintaining the same response time. 
See  the detailed example in Section~\ref{sec:homoarray}.  
Once the opportunity for parallelism has been detected,
additional performance can be gained if the parallel service times can also be reduced. 
See Corollary~\ref{cor:reverse}. 
In Section~\ref{sec:hetero}, 
the generalization of Theorem \eqref{thm:homoq} to the latency optimization of heterogenous parallel arrays is 
presented in Theorem \eqref{thm:heteroq}.

\section{Homogeneous Queues}   \label{sec:homo}
In this section, we present the theorem behind the slightly provocative title that pertains to any 
open network of parallel queues with identical mean service times.

\subsection{Main theorem}
\begin{figure}[!ht]
    \centering
    \begin{subfigure}[c]{0.4\textwidth}
    	\centering
        \includegraphics[scale=0.5]{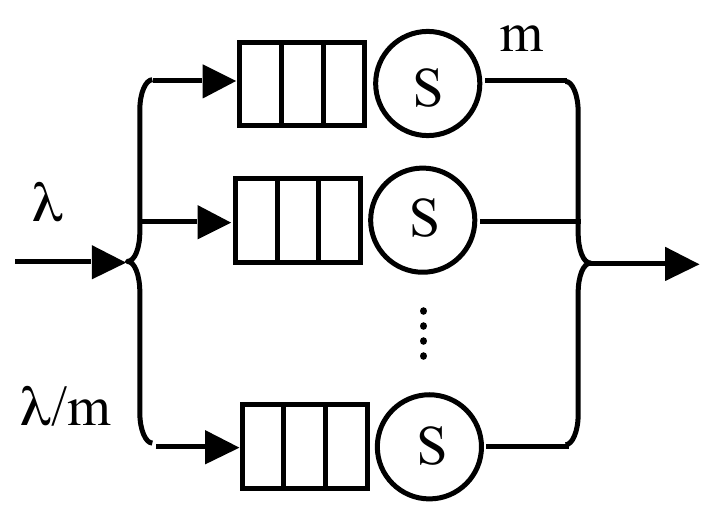}
         \vspace{8pt}
        \caption{Parallel queues $1,2,\ldots,m$, each with a common average service time $S$ but with scaled local arrival rate $\lambda/m$} 
         \label{fig:para}
    \end{subfigure}
    \hspace{0.25in}
    \begin{subfigure}[t]{0.5\textwidth}
    	\centering
        \includegraphics[scale=1.0]{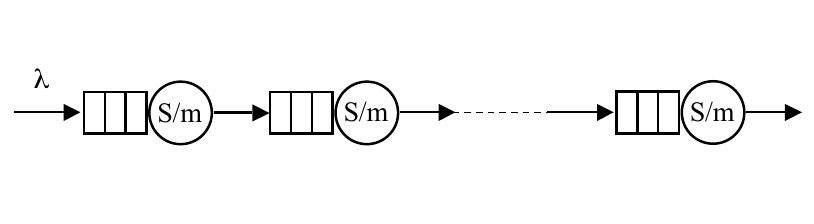}
          \vspace{1pt}
        \caption{Tandem queues $1,2,\ldots,m$,  each with a common global arrival rate $\lambda$ but with a locally scaled average service time $S/m$}    
        \label{fig:serial}
    \end{subfigure}
    \caption{Open parallel and tandem queueing networks with homogeneous service times}. \label{fig:theorem}
\end{figure}

We show that the response-time performance of the queueing network Fig.~\ref{fig:para} and Fig.~\ref{fig:serial} are identical. 

\begin{thm}  \label{thm:homoq}
Each of the open queueing networks in Fig.~\ref{fig:theorem} receive aggregate arrivals due to a Poisson process with rate $\lambda$. 
The mean time spent in Figs.~\ref{fig:para} and~\ref{fig:serial} is the same.
\end{thm}

\begin{prf}
We assume, without loss of generality, that individual queues are M/M/1~\cite{kleinrock}. 
In  Fig.~\ref{fig:para}, the aggregate arrival rate $\lambda$ is split equally $m$-ways each of the queues in the parallel network.  
The arrival rate into any of those queues is thus $\lambda/m$, and the mean time spent in any one of those queues is given by
\begin{equation}
R_{para} = \dfrac{S}{1 - (\lambda/m) \, S}  \label{eqn:para}
\end{equation}
In other words, since all $m$ queues have equal weight, 
the mean residence time is the same as that for a single M/M/1 queue with arrival rate $\lambda/m$.
Referring to Fig.~\ref{fig:serial},  
the mean time spent in the serial queueing network is given by
\begin{equation}
R_{serial}  = \sum_{k=1}^m \dfrac{S/m}{1 - \lambda \, (S/m)} = m \; \bigg(  \dfrac{S/m}{1 - \lambda \, (S/m)}  \bigg)   \label{eqn:msum}
\end{equation}
which simplifies to 
\begin{equation}
R_{serial}  = \dfrac{S}{1 - \lambda \, (S/m)}   \label{eqn:serial}
\end{equation}
Since the resource utilization $\rho$ at each service facility is given by $\rho = (\lambda S) / m$, 
the denominators of \eqref{eqn:para} and \eqref{eqn:serial} are identical  
and therefore, $R_{para} = R_{serial}$.  \qed
\end{prf}

\begin{rem}[Tandem vs. Feedback]
In view of the multiplicative factor of $m$ in \eqref{eqn:msum}, could the tandem queues in Fig.~\ref{fig:serial}  be replaced by a 
single feedback queue~\cite[Ch. 3]{ppa}  with $V=m$ visits? 
Each of the tandem queues only get $V=1$ visit. 
Since the queues are identical,  $R_{serial} = R_1 + R_2 + \cdots + R_m = m \, R$, 
i.e., $m$ scales the residence time $R$.
For a feedback queue with $V=m$ visits,  however, it is the service time that is scaled by $V$
 to produce the service demand $D = V (S/m) = S$.  
Clearly, the residence time  $R_{feedack} = D / (1-\lambda D)$ cannot be equivalent to  $R_{serial}$ in \eqref{eqn:serial}. 
\end{rem}

\subsection{Homogeneous array}  \label{sec:homoarray}
In this section we consider how Theorem~\ref{thm:homoq}  is applied in the context of the 
PDQ analytic queueing solver~\cite{ppa,pdq}. 
In general, 
when using queueing network solvers that employ either analytic or simulation solution techniques, 
there are two approaches for defining a parallel queues like Fig.~\ref{fig:para}.
\begin{description}
\item[Method A:]  
	Since all of the queues in Fig.~\ref{fig:para} are homogenous, any one of them is representative of the others.
	Thus, we only need evaluate a single queue instance. 
	Care needs to be taken that the representative queue only receives the fractional arrival rate $\lambda/m$.  
\item[Method B:]  
	Treat  Fig.~\ref{fig:para} as a  queueing subnetwork that receives the aggregate arrival rate $\lambda$.  
	Each of the $m$ queues can be enumerated separately. 
	The splitting of arrivals within the subnetwork is expressed through the fractional service times $S/m$ 
	by applying Theorem~\ref{thm:homoq}. 
\end{description}

Method A can present programming complications when the parallel queues constitute a subnetwork within a  larger queueing network.  
It necessitates enforcing a consistent distinction between $\lambda$ and $\lambda/m$ in all the right places throughout the global network.
In addition, a lack of explicit instance-naming in the parallel subnetwork can obscure identification of  performance metrics in the reported solution.

Method B is generally safer from a programming standpoint.
The aggregate arrivals only need to be defined once through a global variable $\lambda$, with the proviso that 
each {\em parallel} queue-instance is parameterized by a service time $S/m$ rather than $S$.   
Logically, this is tantamount to a single arrival traversing all the parallel queues in Fig.~\ref{fig:para},  thereby incurring a service time $S$, 
as required.

{\footnotesize  \rm
\begin{lstlisting}[frame=lines, label=code:loop.r, language=R, caption=Explicit parallel queues in the R version of PDQ] 
pdq::CreateOpen(requests, arrivrate)

for (k in 1:pqueues) {
  qname[k] <- sprintf("ParaQ%d", k)
  pdq::CreateNode(qname[k], CEN, FCFS)
  pdq::SetDemand(qname[k], requests, servtime / pqueues)
}
\end{lstlisting}
\normalsize
}

{\footnotesize \rm
\begin{lstlisting}[frame=lines, label=code:loop.rpt, language=bash, caption=Partial PDQ report from listing~\ref{code:loop.r}] 
Metric          Resource     Work              Value   Unit
------          --------     ----              -----   ----
Capacity        ParaQ1       Requests              1   Servers
Throughput      ParaQ1       Requests         0.5000   Requests/Sec
Utilization     ParaQ1       Requests        12.5000   Percent
Queue length    ParaQ1       Requests         0.1429   Requests
. . .		. . .	     . . .	     . . .	. . .
Capacity        ParaQ2       Requests              1   Servers
. . .		. . .	     . . .	     . . .	. . .
Capacity        ParaQ3       Requests              1   Servers
. . .		. . .	     . . .	     . . .	. . .
Capacity        ParaQ4       Requests              1   Servers
. . .		. . .	     . . .	     . . .	. . .
\end{lstlisting}
\normalsize
}

\begin{exam}[Method B in PDQ]
An explicit parameterization using Method B  is shown in Listing~\ref{code:loop.r}---a PDQ model expressed using the R language.
The global variable {\tt arrivrate} corresponds to $\lambda$ as an argument in the function {\tt CreateOpen()}. 
Some select solutions, taken from the corresponding PDQ report, are shown in Listing~\ref{code:loop.rpt}.
The distinct names used to enumerated each of the queues appear in the second column. 
\end{exam}

\begin{cor}   \label{cor:reverse}
Given $m$ queues in tandem with service time $S/m$ and residence time $R_{serial}$,  
reconfiguring them as $m$ parallel queues, while maintaining the serial service time $S/m$, 
reduces the mean parallel response time by a factor of $m$, i.e., $R_{para} = R_{serial} \, / \, m$.
\end{cor}

\begin{exam}[Cloud Application]  \label{ex:cloud}
As alluded to in Section~\ref{sec:intro}, 
Theorem~\ref{thm:homoq} was used in reverse order so as to resolve the PDQ  model of a cloud-based application.
Briefly, some three hundred homogeneous sequential queues, each with $S_{serial} =1$ millisecond, initially had to be incorporated into the PDQ model 
so that the predicted mean response time of $R=300$ milliseconds calibrated with the measured application latency of 3 seconds.
An outstanding question was, what did so many additional queues represent in the actual application? 
Two distinct hypotheses arose:
\begin{enumerate}[(i)]
\item Repetitive or sequential polling of certain resources (cf. Fig.~\ref{fig:serial})
\item Unidentified parallelism in the application  (cf.  Fig.~\ref{fig:para})
\end{enumerate}

\noindent
Based on the original performance data, the polling interpretation seemed the most plausible but could not be easily validated. 
Later performance measurements, however, made it clear that the additional queues were actually associated with parallelism due to the
threaded nature of the application.
Moreover, and consistent with Theorem~\ref{thm:homoq}, $S_{para} = 300 \times S_{serial}$: 
the service time in a parallel queue was indeed three hundred times longer than 
the service time in any of the tandem queues. Full details can be found in~\cite{aws}.
\end{exam}

\subsection{Theorem genesis} 
Having established Theorem~\ref{thm:homoq}, an important question remains: Why has this simple theorem not been disucssed previously in the literature? 
The answer lies in an unanticipated quirk of PDQ (and possibly similar tools) related to how  queueing networks are defined. 
In particular, PDQ has no convenient way to define a parallel subnetwork using Method B.
Parallelism can be expressed in PDQ but it has to be accomplished in a more indirect way than one would expect. 

Method A is often sufficient 
for simple models where the entire queueing network corresponds to Fig.~\ref{fig:para}, 
For example, it is the simplest way to resolve the classic performance question that compares the performance of 
\begin{enumerate*}[(i)] 
\item an $m$-speed single processor, 
\item an $m$-way multicore, and \item an $m$-node cluster. 
\end{enumerate*}
Equation~\eqref{eqn:para} computes the residence time of the $m$-node cluster  
(see~\cite[\S 5.1]{kleinrock2} and~\cite[p.78]{ppa}).

The difficulty arises when the parallel queues belong to a subnetwork within a larger queueing network. 
The PDQ model~\cite[p.195]{ppa} of a 
\href{https://en.wikipedia.org/wiki/DBC_1012}{Teradata DBC 10/12} database cluster machine~\cite{cmg93} represents such a situation. 
Similar to Listing~\ref{code:loop.r}, the arrival rate $\lambda$  into the entire database cluster is defined globally via the 
PDQ function {\tt CreateOpen()}. 
However,  when it comes to the partial arrival rate $\lambda/m$ seen by each local parallel queue, there is no way 
to rescale $\lambda$ using the {\tt SetDemand()} function in PDQ. 

The compromise in PDQ is to define a single $\lambda$ but, 
create $m$ separate queues with each having a rescaled service time $S/m$.
Simultaneously, each queue can be assigned a distinct node-name for later identification in the PDQ Report.
All of this can be accomplished most simply using the loop construct in Listing~\ref{code:loop.r}. 
Remarkably, this procedure is precisely a programmatic representation of Fig.~\ref{fig:serial}. 

It is only within this specific programming context of PDQ that Theorem~\ref{thm:homoq} emerged, and 
 this unique circumstance most likely accounts for why it has not been discussed elsewhere.

\section{Heterogeneous Queues}   \label{sec:hetero}
In this section, we present the generalization of Theorem~\ref{thm:homoq} where the mean service times are no longer assumed to be identical.

\begin{figure}[!ht]
    \centering
    \begin{subfigure}[c]{0.4\textwidth}
    	\centering
        \includegraphics[scale=0.5]{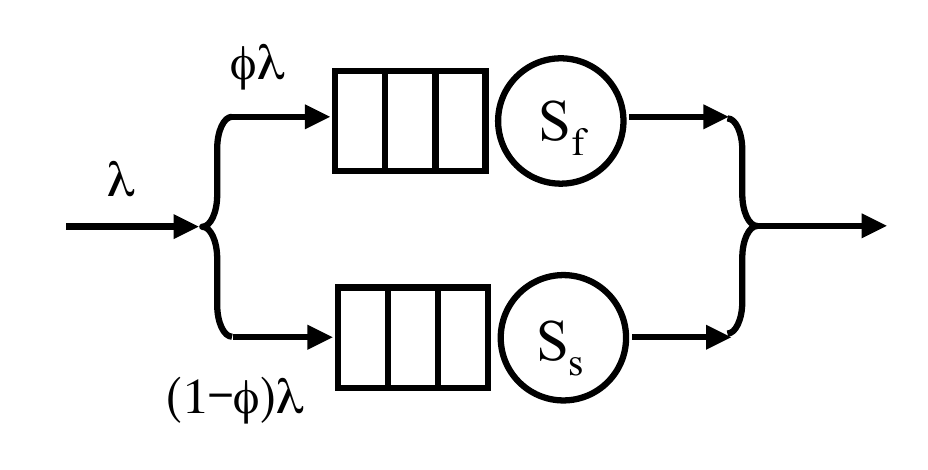}
        \caption{Traffic routing in heterogeneous dual parallel queues} \label{fig:para2}
    \end{subfigure}
    \hspace{0.25in}
    \begin{subfigure}[c]{0.5\textwidth}
    	\centering
        \includegraphics[scale=0.5]{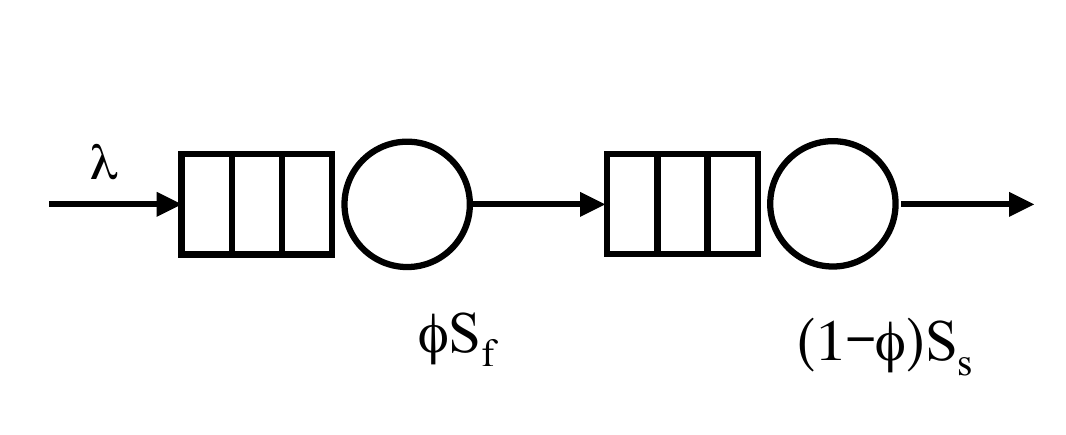}\\[8pt]
        \caption{Tandem queues with heterogeneous effective service times} \label{fig:serial2}
    \end{subfigure}
    \caption{Heterogeneous parallel and tandem queues} \label{fig:dualhet}
\end{figure}

\subsection{Dual heterogeneous disks}
\begin{figure}[!ht]
    \centering
    \begin{subfigure}[c]{0.45\textwidth}
    	\centering
        \includegraphics[scale=0.45]{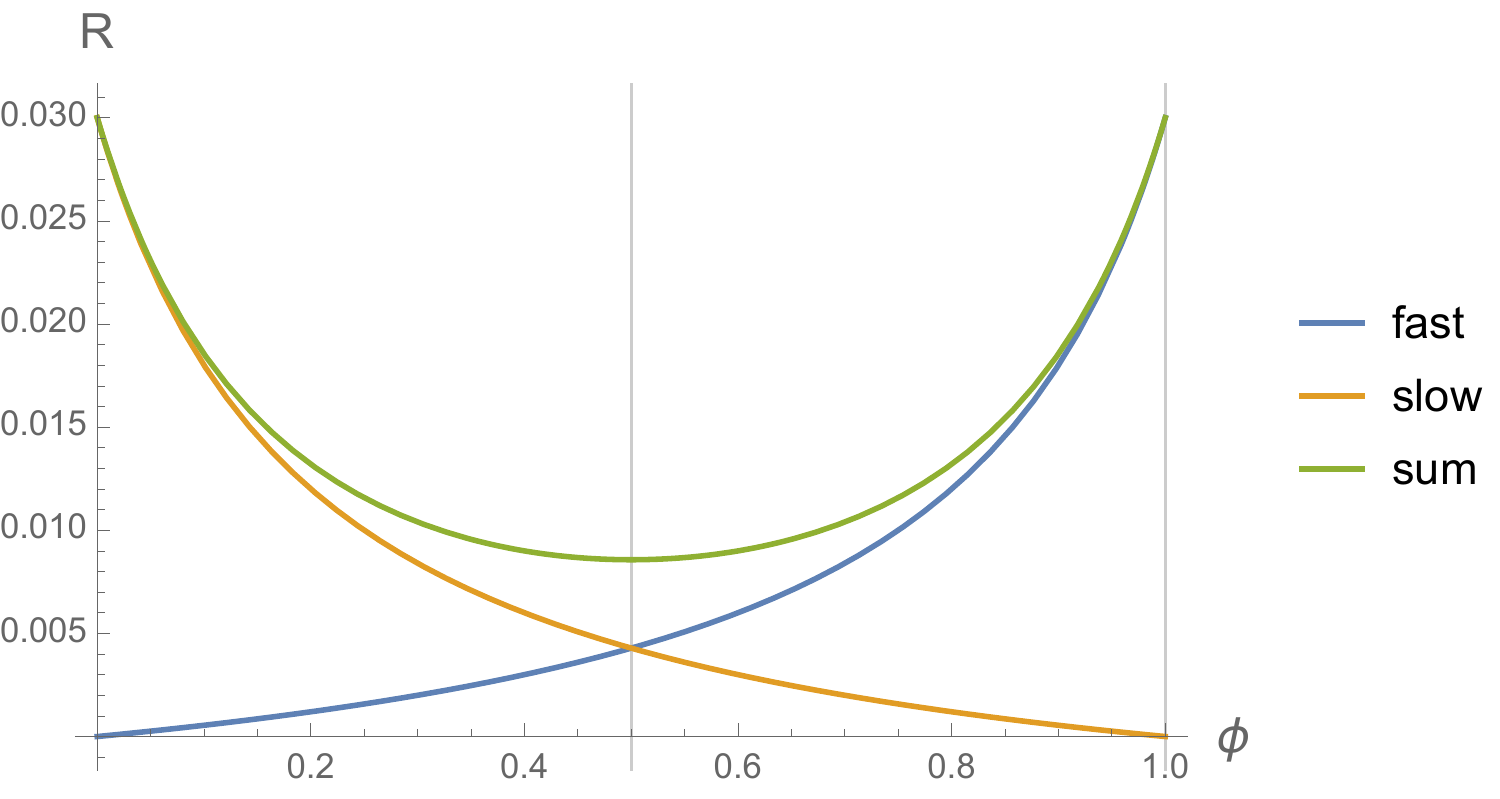}
        \caption{Dual homogeneous queues} \label{fig:homoq}
    \end{subfigure}
    \hspace{0.01in}
    \begin{subfigure}[c]{0.45\textwidth}
    	\centering
        \includegraphics[scale=0.45]{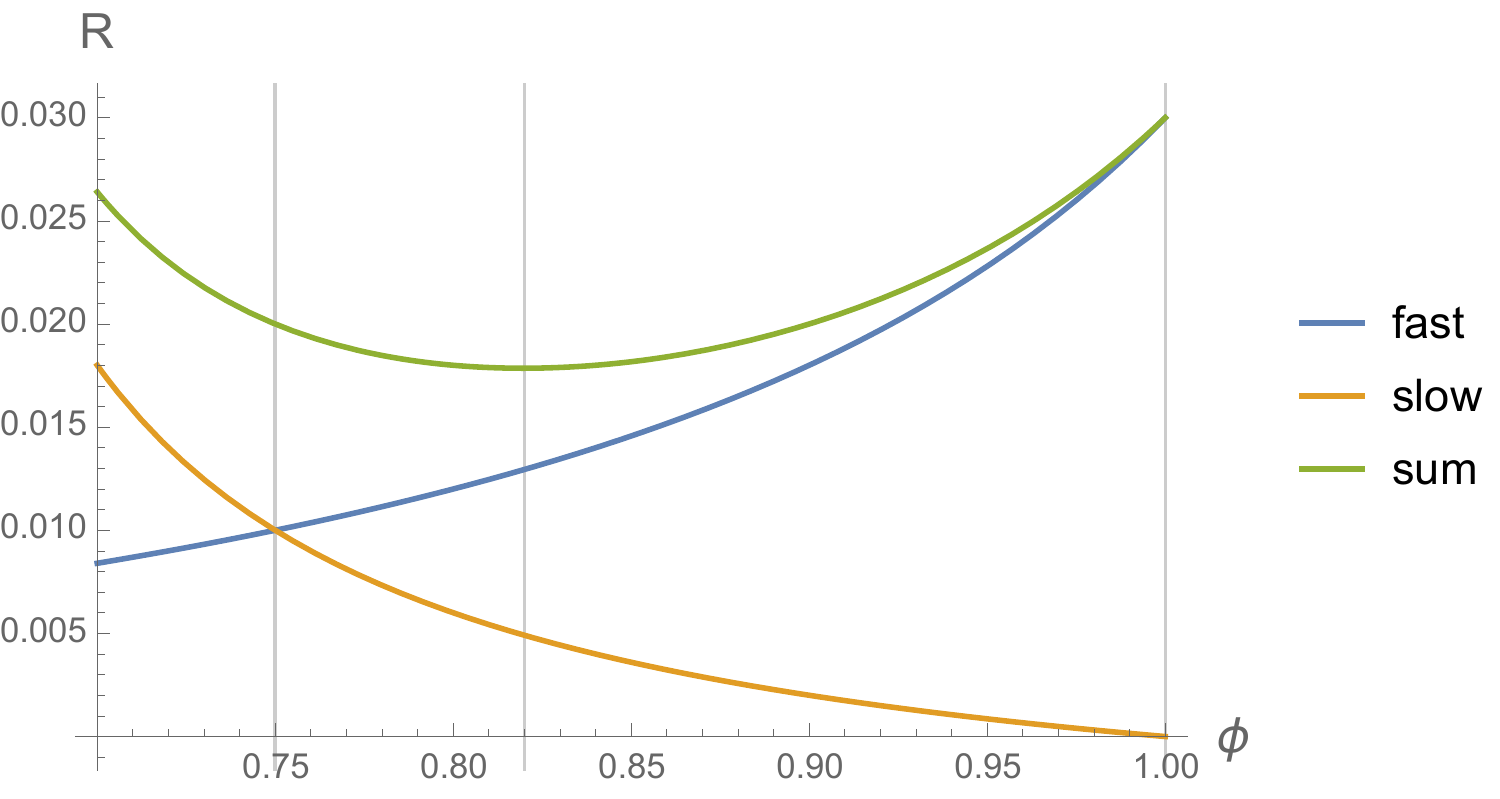}
        \caption{Dual heterogeneous queues} \label{fig:heteroq2}
    \end{subfigure}
    \caption{Homogeneous and heterogeneous response time profiles}\label{fig:queues2}
\end{figure}

If we introduce $0 \leq \phi < 1$ as the fraction of traffic  going to either disk then, in the homogeneous case, $\phi = 0.5$ so as to agree with 
Fig.~\ref{fig:para} when $m=2$.
When one of the dual parallel disks is faster than the other, more traffic can be directed as the faster disk and 
$\phi > 0.5$ and the response time profiles (such as those in Fig.~\ref{fig:heteroq2}) are no longer symmetric.

If the respective fast and slow service times are denoted by $S_f$ and $S_s$, 
the optimal response time of the dual parallel  disks, $R^*_{2}$, is determined by minimizing 
the sum of the residence times in each disk
\begin{equation}
R^*_{2}  = \min_{\phi \in [0,1)} \bigg( \dfrac{\phi \,S_f}{1 - \phi \, \lambda \; S_f} + \dfrac{(1-\phi) \,S_s}{1 - (1-\phi) \, \lambda \; S_s} \bigg) 
\label{eqn:R2}
\end{equation}

\begin{exam}[Dual disk array]  \label{exam:dual}
Let $\lambda = 166.67$ IO/s, $S_f = 0.005$ s, $S_s = 0.015$ s. 
We want to find the fraction of the traffic $\phi$ that should go to the fast disk in order to produce the 
minimum response time $R^*_{2}$.
The derivative of \eqref{eqn:R2} with respect to $f$ is
\begin{equation}
R^\prime_{2}  = \dfrac{\phi  \lambda S_f^2}{(1 - \phi \lambda S_f )^2} + 
\dfrac{S_f}{1 - \phi \lambda S_f } - 
\dfrac{(1 - \phi) \lambda S_s^2 }{(1 - (1 - \phi) \lambda S_s )^2} - 
\dfrac{S_s}{1 - (1 - \phi) \lambda S_s }
\end{equation}
Solving $R^\prime_{2}(\phi) = 0$, yields $\phi = 0.819612$ and $R^*_{2}  = 0.017857$ s (Fig.~\ref{fig:heteroq2}). 
In other words, 82\% of the IO traffic should go to the fast disk in order to minimize the array response time. 
The corresponding symmetric response-time profiles are shown in Fig.~\ref{fig:homoq}.
\end{exam}

\subsection{Heterogeneous array} 

\begin{thm}   \label{thm:heteroq}
An array of $k=1,2,\ldots,m$ parallel queues receives a Poisson arrival stream with aggregate rate $\lambda$.
Each queue has a different service time: $S_1 \leq S_2 \leq \ldots \leq S_m$ with $S_k \in \mathbb{R}^+$.  
The optimal mean response-time is determined by 
\begin{equation}
R^*_{m}  = \min_{\phi_1 + \phi_2 + \cdots + \phi_k = 1} \bigg(  \sum_{k=1}^m \; \dfrac{\phi_k \,S_k}{1 - \phi_k \, \lambda \; S_k} \bigg)  \label{eqn:mdisks}
\end{equation}
where  $\phi_1 \geq \phi_2  \geq \ldots \geq~\phi_m$ are the corresponding routing probabilities $\phi_k \in [0,1)$.
\end{thm}

\begin{prf}
Numerical generalization of \eqref{eqn:R2}.
\end{prf}

\begin{exam}[Quad disk array]
As in example~\ref{exam:dual},  $\lambda = 166.67$ IO/s, but now spread across $m=4$ heterogeneous
disks.  
Solving \eqref{eqn:mdisks} numerically yields the results in Table~\ref{tab:mdisks}.
\begin{table}[htp]
\caption{Numerical optimization of  \eqref{eqn:mdisks} for $m=4$} \label{tab:mdisks}
\centering
\begin{tabular}{c | c | c | c}
Service time & Parameter & Mathematica & R \\
\hline
$S_1 = 0.005$ & $\phi_1$  & 0.734399 & 0.73442474\\
$S_2 = 0.015$ & $\phi_2$  & 0.131191 & 0.13118558\\
$S_3 = 0.020$ &$\phi_3$  &  0.067205 & 0.06719483\\
$S_4 = 0.020$ & $\phi_4$  & 0.067205 & 0.06719483\\
\hline
 & $R^*_{4}$ &0.016524 & 0.01585675\\
\cline{2-4} 
\end{tabular}
\end{table}

\noindent
As expected, the calculated routing weights $\phi_3$ and $\phi_4$ are identical.
\end{exam}

\section{Conclusion} 
P = FS or {\em parallel is just fast serial} is a valuable principle for solving certain types of performance problems. 
Most commonly, it is likely to be used in the context of load-balancing storage arrays, although it is completely generalizable to any type of computational resources, e.g., the cloud-based application described in Example~\ref{ex:cloud}. 

Theorem~\ref{thm:homoq} pertains to a parallel queueing array where the workload is distributed equally across each of the queueing facilities 
due to the mean service times being identical.  
This kind of parallel arrangement can be solved analytically using a queueing analyzer such as PDQ. 
Indeed, as far as we can ascertain,  this seems to be genesis of the original theorem.

In Theorem~\ref{thm:heteroq}, 
we generalized P = FS to parallel queueing arrays where the  mean service times are not identical.  
Heterogeneous load-balancing is an optimization problem that can only be solved numerically. 
Example applications were described in Section~\ref{sec:hetero}.



\end{document}